\documentclass{PoS}
\usepackage{axodraw}
\usepackage{slashed}

\title{Analytic results for two-loop Yang-Mills}

\ShortTitle{Analytic results for two-loop Yang-Mills}

\author{\speaker{David C. Dunbar},\  John H Godwin,\ Guy R. Jehu\  and Warren B. Perkins\\
College of Science, \\
Swansea University, \\
Swansea, SA2 8PP, UK      \\
        E-mail: \email{d.c.dunbar@swan.ac.uk}}

\abstract{Recent Developments in computing very specific helicity amplitudes in two loop QCD are presented. The techniques focus upon the singular structure of the amplitude rather than on a diagramatic and integration approach.}

\FullConference{13th International Symposium on Radiative Corrections\\
		24-29 September, 2017\\
		St. Gilgen, Austria}

\begin{document}

\def\Fcc{P^{(2)}}
\def\spa#1.#2{\left\langle#1\,#2\right\rangle}
\def\spb#1.#2{\left[#1\,#2\right]}
\def\la{\langle}
\def\ra{\rangle}
\def\notag{\nonumber}
\def\ki{a}
\def\kj{b}
\def\kk{c}

\def\Li{\mathop{\hbox{\rm Li}}\nolimits}
\def\Log{{\rm Log}}
\def\tr{{\rm tr}}
\def\ksl{\slashed{k}}

\def\Psl{\slashed{K}}

\section{Introduction}

There has been excellent progress in computing the matrix elements for $2\longrightarrow2$ NNLO processes which together with work
on factorisation has led to robust predictions for many processes.  For higher point matrix elements there has been very limited 
progress except in highly symmetric, particularly supersymmetric, theories and indeed the only for four points are the two-loop QCD amplitude 
known for all helicities~\cite{Glover:2001af,Bern:2002tk,Abreu:2017xsl}.  Beyond four point 
the five-point all-plus amplitude which  recently constructed using generalised unitarity 
techniques~\cite{Badger:2013gxa}
followed by integration~\cite{Gehrmann:2015bfy}.   
In this talk, it was shown how, for this 
very specific helicity configuration, the singular structure of the amplitude can be used to determine the two loop-amplitude.  
Specifically we have re-computed the five-point case~\cite{Dunbar:2016aux} and obtained results for the six and seven-point 
all-plus amplitudes~\cite{Dunbar:2016gjb,Dunbar:2017nfy}. 

The all-plus helicity amplitude at leading colour may be written
\begin{eqnarray}
{\mathcal A}_{n}(1^+, 2^+,..., n^+) |_{\rm leading \; color}=& g^{n-2}  \sum_{L \ge 1} \left( g^2 N_c c_{\Gamma}\right)^L  
\times \sum_{\sigma \in S_{n}/Z_{n}} {\rm tr}(T^{a_{\sigma(1)}} T^{a_{\sigma(2)}} 
T^{a_{\sigma(3)}} \cdots  T^{a_{\sigma(n)}}) 
\nonumber  \\ &
\times A^{(L)}_{n}(\sigma(1)^{+},  \sigma(2)^{+} ,..., \sigma(n)^{+})\,
\label{eq:fullamp}
\end{eqnarray}
and it is $A^{(2)}_n(1^+,2^+,\cdots,n^+)$ which is the subject of this talk. 
This helicity amplitude vanishes at tree level and consequently has a purely rational one-loop expression to order $\epsilon$
given by~\cite{Bern:1993qk} 
\begin{eqnarray}
A^{(1)}_n(1^+,2^+,\cdots,n^+)=-{i\over 3}\sum_{1\leq k_1<k_2<k_3<k_4\leq n} 
{\spa{k_1}.{k_2} \spb{k_2}.{k_3}\spa{k_3}.{k_4}\spb{k_4}.{k_1} \over \spa{1}.2\spa{2}.3 \cdots\spa{n}.1}  
+O(\epsilon) \; . 
\end{eqnarray}
while
the all-$\epsilon$ forms of the one-loop amplitudes are given in terms of higher dimensional scalar integrals~\cite{Bern:1996ja}.
and for $n\leq 6$ are~\cite{Bern:1996ja}
\def\eps{\epsilon}
$$
A_{4}^{(1)}(1^+,2^+,3^+,4^+) =
{ 2i \eps(1-\eps)\over  \spa1.2\spa2.3\spa3.4\spa4.1}
\times s_{12}s_{23} I_4^{D=8-2\eps} \,,
$$
\begin{eqnarray}
A_{5}^{(1)}(1^+,2^+,3^+,4^+,5^+) &=&
{ i \eps(1-\eps)\over  \spa1.2\spa2.3\spa3.4\spa4.5\spa5.1}
\notag \\
&\times &
\Bigl[
s_{23}s_{34} I_4^{(1),D=8-2\eps}
+s_{34}s_{45} I_4^{(2),D=8-2\eps}
+s_{45}s_{51} I_4^{(3),D=8-2\eps}
\notag \\
&+ & s_{51}s_{12} I_4^{(4),D=8-2\eps}
+s_{12}s_{23} I_4^{(5),D=8-2\eps}
+(4-2\eps) { \varepsilon (1,2,3,4) }I_5^{D=10-2\eps}
\Bigr] \,,
\notag
\end{eqnarray}
\begin{eqnarray}
 A_{6}^{(1)}(1^+, 2^+, 3^+, &   4^+ & , 5^+, 6^+) =
 {i \eps(1-\eps)\over \spa1.2 \spa2.3 \spa3.4 \spa4.5 \spa 5.6 \spa6.1}
 \frac{1}{2} \biggl[
  \notag \\
&  & 
- \hskip -.3 cm \sum_{1\leq i_1<i_2 \leq 6} \hskip -.2 cm
\tr[\ksl_{i_1} \Psl_{i_1+1, i_2-1}
\ksl_{i_2} \Psl_{i_2+1,i_1-1}] I_{4}^{(i_1,i_2),D=8-2\eps}
+  (4-2\eps)\, \tr[123456] \, I_6^{D=10-2\eps} \notag \notag \\
&  & \hskip 2 cm
+  (4-2\eps) \sum_{i=1}^6 \varepsilon(i+1, i+2, i+3, i+4)
I_{5}^{(i),D=10-2\eps} 
\biggr] \,, 
\end{eqnarray}
where $I_m^{(i),D}$ denotes the $D$ dimensional scalar integral obtained by removing the
loop propagator between legs $i-1$ and $i$ from the $(m+1)$-point
scalar integral etc.~\cite{Bern:1993kr},  $\Psl_{a,b}\equiv \sum_{i=a}^b \slashed{k}_i$ and
$\varepsilon(a,b,c,d)=\spb{a}.b\spa{b}.c\spb{c}.d\spa{d}.a-\spa{a}.b\spb{b}.c\spa{c}.d\spb{d}.a$.

\section{Techniques}

We will attempt to compute the amplitude purely from the singularities much in the tradition of $S$-matrix theory~\cite{Eden}. 
Specifically we will consider

\begin{center}
$\bullet$ IR and UV singular structure  under regularisation

$\bullet$  Unitarity \hskip 6.2 truecm $\null$

$\bullet$ Factorisation \hskip 5.6 truecm $\null$
\end{center}

\vskip 0.3 truecm 

\noindent
{\bf Regularisation structure.}
\noindent
The IR and UV behaviours of the two-loop amplitude in dimensional regularisation are known~\cite{Catani:1998bh} and for this amplitude 
motivates a partition:
\begin{eqnarray}
\label{definitionremainder}
A^{(2)}_{n} (1^+, 2^+,..., n^+)=& A^{(1)}_{n}(1^+, 2^+,..., n^+)
I_n^{(2)}
+  \; F^{(2)}_{n}  +   {\mathcal O}(\epsilon)\, .
\end{eqnarray}
where
\begin{equation}
I_n^{(2)}= \left[ - \sum_{i=1}^{n} \frac{1}{\epsilon^2} \left(\frac{\mu^2}{-s_{i,i+1}}\right)^{\epsilon}  \right] 
\end{equation}
In this equation $A^{(1)}_{n}$ is the all-$\epsilon$ form of 
the one-loop amplitude.   
There are no $\epsilon^{-1}$ terms in this expression (outside of $I_n$) 
although the amplitude has both a UV divergence and a collinear IR divergence~\cite{Kunszt:1994np}. However 
since the tree amplitude vanish both are proportional to $n$ and cancel leaving only the 
infinities within $I_n^{(2)}$ which are the soft IR singular terms. 
The finite remainder function $F_n^{(2)}$ can be split into polylogarithmic and rational pieces,
\begin{equation}
F_n^{(2)} = \Fcc_n+R_n^{(2)}\; .
\end{equation}

\vskip 0.3 truecm
\noindent
{\bf Unitarity.}
$D$-dimensional unitarity techniques can be used to generate the integrands~\cite{Badger:2013gxa} for the five-point amplitude which can then be integrated to give the result~\cite{Gehrmann:2015bfy}.   However the organisation of the amplitude in the previous section allows us to obtain the finite polylogarithms using four-dimensional unitarity~\cite{Bern:1994zx,Bern:1994cg} where the cuts are evaluated in four dimension with the corresponding simplifications.  
With this simplification the all-plus one-loop amplitude effectively becomes an additional on-shell vertex and the two-loop cuts effectively become one-loop cuts with a single insertion of this vertex.  The non-vanishing four dimensional cuts are shown in fig.~1. 

  \begin{figure}[h]
\centerline{
    \begin{picture}(170,150)(-150,-40) 
    \Text(-24,92)[c]{$a)$}   
     \Line( 0, -24)( 0,60)
     \Text(0,-32)[c]{$_+$}
     \Text(-14,75)[c]{$_+$}
     \Text(74,-15)[c]{$_+$}
     \Text(-32,0)[c]{$_+$}
     \Vertex(-6,-18){2}
     \Vertex(-18,-6){2}
     \Vertex(-13,-13){2}
     \Line( 0,60)(60,60)
     \Line(60,60)(60, 0)
     \Line(60, 0)( -24, 0)
     \Line( 0,60)(-10,70)
     \Line(60,60)(60,84)
     \Line(60,60)(84,60)
     \Text(60,92)[c]{$_+$}
     \Text(92,60)[c]{$_+$}
     \Vertex(66,78){2}
     \Vertex(78,66){2}
     \Vertex(73,73){2}
     \Line(60, 0)(70,-10)
     \DashLine(52,30)(68,30){2}
     \DashLine(-8,30)(8,30){2}
     \DashLine(30,8)(30,-8){2}
     \DashLine(30,52)(30,68){2}
      \Text(-5,25)[c]{$_-$}
      \Text(-5,35)[c]{$_+$}
      \Text(25,-5)[c]{$_-$}
      \Text(35,-5)[c]{$_+$}
      \Text(65,25)[c]{$_-$}
      \Text(65,35)[c]{$_+$}
      \Text(25,65)[c]{$_-$}
      \Text(35,65)[c]{$_+$}
   \CCirc(60,60){5}{Black}{Purple} 
    \end{picture} 
    \begin{picture}(170,150)(-110,-40)    
     \Text(-24,92)[c]{$b)$} 
     \Line( 0, 0)(30,60)
     \Line(60, 0)(30,60)
     \Line(60, 0)( 0, 0)
     \Line( 0, 0)(-24,0)
     \Line( 0, 0)(0,-24)
     \Text(0,-32)[c]{$_+$}
     \Text(-32,0)[c]{$_+$}
     \Vertex(-6,-18){2}
     \Vertex(-18,-6){2}
     \Vertex(-13,-13){2}
     \Line(30,60)(50,80)
     \Line(30,60)(10,80)
     \Text(55,85)[c]{$_+$}
     \Text(5,85)[c]{$_+$}
     \Text(74,-15)[c]{$_+$}
     \Vertex(40,75){2}
     \Vertex(20,75){2}
     \Vertex(30,78){2}
     \Line(60, 0)(70,-10)
     \DashLine(39,25)(53,33){2}
     \DashLine(21,25)(7,33){2}
     \DashLine(30,8)(30,-8){2}
     \Text(14,37)[c]{$_+$}
     \Text(8,25)[c]{$_-$}
     \Text(46,37)[c]{$_+$}
     \Text(52,25)[c]{$_-$}
     \Text(25,-5)[c]{$_-$}
     \Text(35,-5)[c]{$_+$}
    \CCirc(30,60){5}{Black}{Purple} 
    \end{picture} 
   \begin{picture}(170,150)(-40,-40)    
     \Text(-24,92)[c]{$c)$} 
     \Line( 0, 0)(30,60)
     \Line(60, 0)(30,60)
     \Line(60, 0)( 0, 0)
     \Line( 0, 0)(-10,-10)
     \Line(30,60)(50,80)
     \Line(30,60)(10,80)
     \Vertex(40,75){2}
     \Vertex(20,75){2}
     \Vertex(30,78){2}
     \Line(60, 0)(70,-10)
     \Text(74,-15)[c]{$_+$}
     \Text(-14,-15)[c]{$_+$}
     \DashLine(39,25)(53,33){2}
     \DashLine(21,25)(7,33){2}
     \DashLine(30,8)(30,-8){2}
     \Text(14,37)[c]{$_+$}
     \Text(8,25)[c]{$_-$}
     \Text(46,37)[c]{$_+$}
     \Text(52,25)[c]{$_-$}
     \Text(25,-5)[c]{$_-$}
     \Text(35,-5)[c]{$_+$}
     \Text(25,5)[c]{$_+$}
     \Text(35,5)[c]{$_-$}
   \CCirc(30,60){5}{Black}{Purple} 
    \end{picture} 
    \begin{picture}(170,150)( 30,-40)  
     \Text(-24,92)[c]{$d)$} 
     \CArc(30,30)(20,0,360)
     \Line(10,30)( -12,45)
     \Line(10,30)( -12,15)
     \Text(-15,48)[c]{$_+$}
     \Text(-15,12)[c]{$_+$}
     \Vertex(-8,38){2}
     \Vertex(-10,30){2}
     \Vertex(-8,22){2}
     \Line(50,30)(72,45)
     \Line(50,30)(72,15)
     \Text(75,48)[c]{$_+$}
     \Text(75,12)[c]{$_+$}
     \Vertex(68,38){2}
     \Vertex(70,30){2}
     \Vertex(68,22){2}
    \CCirc(50,30){5}{Black}{Purple}
     \DashLine(30,43)(30,59){2}
     \DashLine(30,17)(30,1){2}
     \Text(25,55)[c]{$_-$}
     \Text(25,5)[c]{$_-$}
     \Text(35,55)[c]{$_+$}
     \Text(35,5)[c]{$_+$}    
 \end{picture} 
    }
    \caption{Four dimensional cuts of the two-loop all-plus amplitude involving an all-plus one-loop vertex 
    (indicated by $\bullet\;$ )} 
    \label{fig:oneloopstyle}
\end{figure}
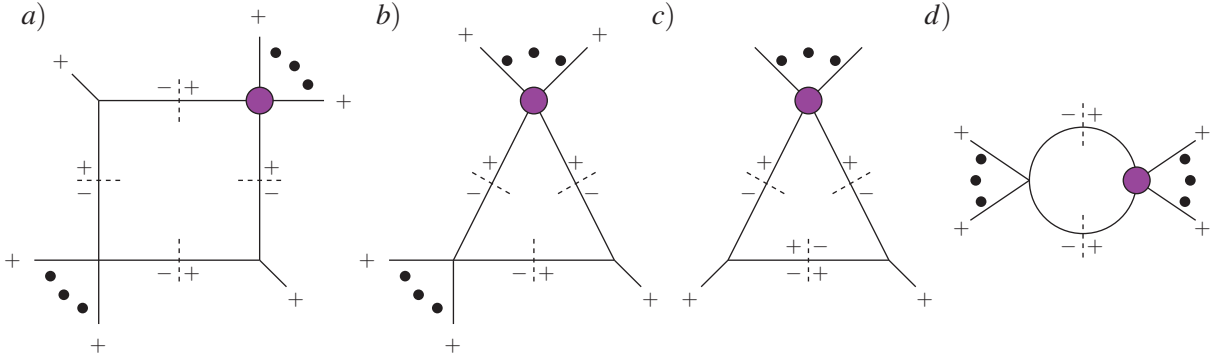

The cuts allow us to determine the coefficients of box and triangle functions to the amplitude. These contain both IR terms and finite polylogarithms. The IR terms combine
overall~\cite{Dunbar:2016cxp}, 
\begin{eqnarray}&
\sum  {\cal C}_{i} I_{4,i}^{\rm 2m} 
\biggl|_{IR}
+\sum  {\cal C}_{i}  I_{3,i}^{2 \rm m}  
+\sum  {\cal C}_{i}  I_{3,i}^{1 \rm m} 
= 
A^{(1),\epsilon^0}_n(1^+,2^+,\cdots, n^+)
\times I_n
\end{eqnarray}
where $A^{(1),\epsilon^0}_n(1^+,2^+,\cdots, n^+)$ is the order $\epsilon^0$ truncation of the one-loop amplitude. 
A key step is to promote the coefficient of these terms to the all-$\epsilon$ form of the one-loop amplitude. This ensures that the two-loop amplitude
has the correct singular structure. 

The remaining parts of the box integral functions become the polylogarithms.
The full expression for $\Fcc_n$ is~\cite{Dunbar:2016cxp} is 
\begin{equation}
P_n^{(2)}   =  -{ i  \over 
3 \spa1.2\spa2.3\spa3.4\cdots \spa{n}.1   }\sum_{i=1}^n  \sum_{r=1}^{n-4} c_{r,i}  F^{2m}_{n:r,i}
\end{equation}
where
\begin{eqnarray}
c_{r,i}= \left( 
\sum_{a<b<c<d \in K_4}  \tr_{-} [abcd]-\sum_{a<b<c \in K_4} \tr_{-}[abc  K_4]  +\sum_{a<b \in K_4 } { \la {i-1} |K_4  a b K_4 | {i+r} \ra  \over \spa{{i-1}}.{{i+r}} }
\right)\; ,
\end{eqnarray}
\begin{equation}
F^{2m}_{n:r,i}=F^{2m}[ t_{i-1}^{[r+1]},t_i^{[r+1]},  t_{i}^{[r]}, t_{i+r+1}^{[n-r-2}] \; ,
\end{equation}
 $t_i^{[r]} =(k_i+k_{i+1}+\cdots +k_{i+r-1})^2$ and
\begin{eqnarray}
F^{2m}[ S,T, K_2^2, K_4^2] = 
&\Li_2[1-\frac{K_2^2}{S}]+
\Li_2[1-\frac{K_2^2}{T}]+
\Li_2[1-\frac{K_4^2}{S}]
\notag \\
+
&
\Li_2[1-\frac{K_4^2}{T}]-
\Li_2[1-\frac{K_2^2K_4^2}{ST}]+
\Log^2(S/T)/2 \; . 
\end{eqnarray}

\vskip 0.5 truecm 
\noindent
{\bf Factorisation.}
The remaining part of the amplitude is the rational $R_n^{(2)}$.  As a rational function we may wish to obtain this via recursion
provided we can control its singularities.
We wish to use complex recursion to determine $R(z)$. Britto-Cachazo-Feng-Witten recursion \cite{Britto:2005fq} exploited the analytic properties of $n$-point tree amplitude under
a complex shift of its external momenta to compute the amplitude. 
The momentum shift
introduces a complex parameter, $z$, whilst  preserving  overall momentum conservation and keeping all external momenta null.  
Possible shifts include the original BCFW shift which acts on two momenta, say  $p_\ki$ and $p_\kj$, by
\begin{equation}
\bar\lambda_{{\ki}}\to \bar\lambda_{\hat{\ki}} =\bar\lambda_\ki - z \bar\lambda_\kj 
,
\lambda_{{\kj}}\to\lambda_{\hat{\kj}} =\lambda_\kj + z \lambda_\ki \; ,
\end{equation}
and
the Risager shift~\cite{Risager:2005vk} which acts on three momenta, say $p_\ki$, $p_\kj$ and $p_\kk$, by shifting $\lambda_\ki$
\begin{eqnarray}
\lambda_\ki\to \lambda_{\hat \ki} = \lambda_\ki\,\, +z\, [\kj\kk] \lambda_\eta \,,
\label{KasperShift}
\end{eqnarray}
and cyclically $\lambda_\kj$ and $\lambda_\kk$. 
In the last case $\lambda_\eta$ must satisfy $\spa{\ki}.{\eta}\neq 0$ etc., but is otherwise unconstrained. 
After applying the shift, the rational quantity of interest is a complex function parametrized by $z$ i.e. $R(z)$. 
If $R(z)$ vanishes at large $\vert z\vert$, the
Cauchy's theorem applied to $R(z)/z$ over a contour at infinity implies
\begin{equation}
R= R(0)= -\sum_{z_j\neq 0} {\rm Res}\Bigl[ {R(z)\over z}\Bigr]\Bigr\vert_{z_j}\; .
 \label{AmpAsResidues2}
\end{equation}
Tree amplitudes have simple poles when a 
shifted propagator vanishes and the corresponding residues  are readily obtained from general factorisation theorems
leading to the BCFW recursion formulae for tree amplitudes~\cite{Britto:2005fq}. 
For the rational part of the two-loop all-plus amplitude the 
BCFW shift generates a shifted quantity that does not vanish at infinity and so cannot be used to reconstruct the amplitude (the one-loop all-plus amplitudes
also behave in this way). However, using the Risager shift~(\ref{KasperShift}) does yield a shifted quantity with the desired asymptotic behaviour. Also loop amplitudes in non-supersymmetric theories may have double poles in complex momenta. Mathematically this is not a problem since if we consider a function with a 
double pole at $z=z_j$  and 
Laurent expansion,
\begin{eqnarray}
R(z) &=& \frac{c_{-2}}{(z-z_{j})^2}+ \frac{c_{-1}}{(z-z_{j})}+\mathcal{O}((z-z_{j})^0)\; , \notag \\
\end{eqnarray}
then the required residue is 
\begin{equation}
{\rm Res}\Bigl[ {R(z)\over z}\Bigr]\Bigr\vert_{z_j}  = 
-\frac{c_{-2}}{z_{j}^2}+ \frac{c_{-1}}{z_{j}}
\end{equation}
and we can use Cauchy's theorem {\it provided} we know the value of both the leading and sub-leading poles. 
The leading pole can be obtained from factorisation theorems, but,
at this point, there are no general theorems determining the sub-leading pole and we need to determine the 
sub-leading pole for each specific case.

\begin{figure}[h]
\begin{center}
\includegraphics[scale=0.8]{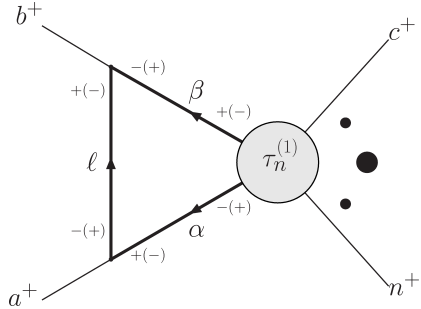}
\end{center}
\caption{Diagram containing the leading and sub-leading poles as $s_{ab}\to 0$. The axial gauge construction permits the off-shell continuation of the internal 
legs.}
\label{fig:axialg}
\end{figure} 

We determine the sub-leading pole by determining the pole in the diagram shown in fig~2 using an axial gauge formalism.
We have used this approach previously to compute one-loop 
amplitudes~\cite{Dunbar:2010xk,Alston:2015gea,Dunbar:2016dgg}.
and labeled this process  {\em augmented recursion}. 
The principal helicity assignment in fig~2 gives the integral
\begin{eqnarray}
 \frac{i}{(2\pi)^D}\int\!\! \frac{d^D\ell}{\ell^2\alpha^2\beta^2} 
\frac{[a|\ell|q\ra[b|\ell|q\ra }{\spa a.q\spa b.q} \frac{\spa \beta.q^2}{\spa\alpha.q^2}\tau_n^{(1)} (\alpha^{-},\beta^{+},c^+,...,n^+) \; .
\label{eq:tauintinitA}
\end{eqnarray}
To determine \ref{eq:tauintinitA} in general we would need to consider $\tau_n^{(1)}$ to be the 
doubly off-shell one-loop current. 
However, as we are only interested in the residue on the $s_{ab}\to 0$ pole, we do not need the full current.
This process is detailed in ref.~\cite{Dunbar:2017nfy}.  The resultant sub-leading pole is quite complex but can be substituted into~\ref{AmpAsResidues2} to yield the unshifted $R(0)$. The initial expression, after combining all factorisation, 
can be simplified into quite compact forms. 
We obtain a form for $R_6^{(2)}$  that is explicitly independent of $q$, has manifest cyclic symmetry and no spurious poles. 
\begin{eqnarray}
R_6^{(2)} = {i\over 9}\sum_{{cyclic perms}}{G_6^1+G_6^2+G_6^3+G_6^4+G_6^5\over \spa 1.2\spa 2.3 \spa 3.4 \spa 4.5 \spa 5.6 \spa 6.1}.
\end{eqnarray}
where
\begin{eqnarray}
G^1_6 &=&   { s_{cd} s_{df} \la f|a K_{abc}|e \ra\over \spa{f}.{e} t_{abc}} + { s_{ac} s_{dc} \la a|f K_{def}|b \ra\over \spa{a}.{b} t_{def}}\; ,
 \nonumber \\
G^2_6 &=&   {\spb{a}.{b} \spb{f}.{e} \over \spa{a}.{b} \spa{f}.{e} }  \spa{a}.{e}^{2} \spa{f}.{b}^2  +  {1 \over 2}{\spb{a}.{f} \spb{c}.{d} \over \spa{a}.{f} \spa{c}.{d} }  \spa{a}.{c}^{2} \spa{d}.{f}^2   \; ,
 \nonumber \\
 G^3_6 &=&   { s_{df} \spa{a}.{f} \spa{c}.{d} \spb{c}.{a} \spb{d}.{f} \over t_{abc}}  
 \nonumber \\
 \;\; , \;\;
G^4_6 &=&   { \la a |b e|f \ra t_{def} \over \spa{a}.{f} } \; \;\; 
\end{eqnarray}
and
\begin{eqnarray}
G^5_6 =   & 2s_{ac}^2+s_{eb}^2 + s_{ab} \left(-3 s_{ac} - 2 s_{ad} +6 s_{ae} +4s_{bc} + s_{bd} +2s_{be} + 4s_{bf}+7 s_{cd} - s_{ce} - s_{de} +3 s_{df}\right) \nonumber  \\ 
&+s_{ac} \left( 2s_{ad} +3s_{ae} -2 s_{bd} - s_{be} + s_{cf} -{5 \over 2}s_{df} \right) + {3 \over 2 }s_{ad}s_{be} \nonumber  \\
&- 8 \spa{b}.{c} \spb{c}.{d} \spa{d}.{e} \spb{e}.{b} +5 \spa{f}.{a} \spb{a}.{c} \spa{c}.{d} \spb{d}.{f}\; ,
\end{eqnarray}
This was confirmed in an subsequent independent calculation~\cite{Badger:2016ozq}.

\section{The Seven-Point Rational Piece}
We have also computed the seven point~$R_7^{(2)}$ ref.~\cite{Dunbar:2017nfy}.
The seven-point rational piece can be calculated in an identical fashion. The seven-point current $\tau_7^{(1)} (\alpha^{-},\beta^{+},c^+,d^+,e^+,f^+,g^+)$
is built from the corresponding seven-point single minus amplitude~\cite{Bern:2005hs} just as the six-point current was built from the six-point amplitude. 
$R_6^{(2)}$ as determined above is also required for recursion. Defining
\begin{eqnarray}
G^1_7  & =  & {\spa g.a\over t_{abc}t_{efg}} \Biggr({\spa c.d\spb e.g[d|K_{abc}|e\ra[a|K_{abc}|e\ra [c|K_{abc}|f\ra\over \spa e.f}-{\spa d.e\spb c.a [d|K_{efg}|c\ra  [g|K_{efg}|c\ra [e|K_{efg}|b\ra  \over \spa b.c}
\notag \\
 &
+  & { \spa e.f \spa c.d  \spb c.a\spb f.g [e|K_{efg}|a\ra [d|K_{efg}| b\ra \over \spa a.b} - {\spa b.c \spa d.e \spb e.g \spb a.b [c|K_{abc}|g\ra [d|K_{abc}| f\ra\over \spa f.g}\Biggr)\; ,
 \notag \\
G^2_7  & =  & {1\over t_{abc}t_{efg}}\, s_{cd}s_{de}\spa g.a[g|K_{efg}K_{abc}|a] \; ,
\notag \\
G^3_7 & =  & {1\over t_{cde}} \Biggr( s_{ce}\Biggr({s_{ef}\la c|K_{ab}K_{fga}|d\ra\over \spa c.d}-{s_{bc}\la e|K_{fg}K_{gab}|d\ra \over \spa d.e}\Biggr)+{\spa e.f \spa b.c \spb f.b[c|K_{cde}|g\ra[e|K_{cde}|a\ra\over \spa g.a}
\notag \\ &+ &{\spa{b}.c [c|K_{cde}|b\ra[e|K_{cde}|a\ra[b|K_{fg}|e\ra\over \spa a.b}
+{ \spa{e}.f [e|K_{cde}|f\ra[c|K_{cde}|g\ra[f|K_{ab}|c\ra\over \spa f.g}\Biggr) \; ,
 \notag \\
G^4_7 &=  & {\spb g.a\over \spa g.a} \spa g.e \spa a.e\Biggr( {\spb d.e\over \spa d.e}\spa d.g \spa d.a+{\spb e.f\over \spa e.f}\spa f.g\spa f.a \Biggr)\ \; ,
 \notag \\
G^5_7 &=  & {1\over t_{cde}}\big( \spb c.e (\spa e.f \spb d.f \la c|K_{ab} K_{fga}|d\ra+\spa b.c \spb d.b \la e|K_{fg}K_{gab}|d\ra  )
\notag \\ & &
\quad\quad\quad\quad\quad\quad\quad+ \spa b.c\spa e.f(2 \spa g.a \spb c.e\spb f.g \spb a.b+  \spb b.f [e|K_{ab}K_{fg}|c]\big) \; ,
 \notag \\
G^6_7 & =  & {1\over\spa g.a} (\la g|fK_{bc}|a\ra t_{efg} -\la a|bK_{ef}|g\ra t_{abc}) 
\notag \\ 
G^7_7  & =  & s_{bf}^2-2s_{ga}^2-3s_{db}s_{df}+4s_{da}s_{dg}-6s_{ac}s_{eg}+7(s_{eb}s_{fc}+s_{ea}s_{gc})+s_{ab}s_{fg}+3s_{fa}s_{gb}
 \notag \\ &  &
  +s_{ce}(s_{cf}+s_{eb}-4(s_{ab}+s_{fg}+s_{ga})+5[d|K_{ga}|d\ra)
  \notag \\ & &
  +4[e|bcf|e\ra-2[f|gab|f\ra+3[g|baf|g\ra+2[g|cea|g\ra  ,
\end{eqnarray}
the full function in this case is
\begin{eqnarray}
R_7^{(2)} = {i\over 9}\sum_{\hbox{cyclic perms}}{G_7^1+G_7^2+G_7^3+G_7^4+G_7^5+G_7^6+G_7^7\over \spa 1.2\spa 2.3 \spa 3.4 \spa 4.5 \spa 5.6 \spa 6.7 \spa 7.1}\;.
\end{eqnarray}
This expression has the full cyclic and flip symmetries  required and has all the correct factorisations and collinear limits. 
It been generated under the assumption that the shifted rational function vanishes at infinity: if this had been unjustified we would not 
have generated a function with the appropriate symmetries. 
This completes the seven-point calculation: the first seven point helicity amplitude obtained in QCD.

\section{Summary and Prospects} 
\cite{Abreu:2017hqn}
We have been able to use the singularity structure of amplitudes to obtain higher-point two-loop QCD amplitudes.  The methods have avoided 
complicated two-loop integrations. Correspomding results have been obtained for two-loop  gravity where the UV 
term may be simple obtained~\cite{Dunbar:2017qxb}.
So far, we have only been able to generate the simplest helicity amplitude by these methods but have obtained these in  compact analytic forms
which complement recent progress in numerical techniques~\cite{Badger:2017jhb,Abreu:2017hqn}
The effort to extend these methods to further helicity configurations is on-going.

\end{document}